# On the Capacity of Reconfigurable Intelligence Surface: the Sparse Channel Case

Chenxi Zhu, *Senior Member, IEEE*

zhucx1@lenovo.com

Lenovo Research

**Abstract:** Reconfigurable intelligent surface (RIS) is an important candidate technology for 6G. We provide an analysis of RIS-assisted MIMO communication in sparse channel typically found in the mmW or THz range. By exploring the sparse property, we maximize the capacity in the singular space of the channel and developed efficient algorithms for SU-MIMO or DL MU-MIMO. We also proved it is more difficult to support high rank transmission in the RIS reflection channel than in the traditional MIMO channel.

## 1. Introduction

With rapid development and deployment of 5G, the wireless research community has turned its attention to the next generation of wireless technology. Recently, through the collective efforts of many institutions, vision for 6G network starts to emerge [1][2]. It has been widely agreed that 6G will further extend the scenarios initiated by 5G (eMBB, URLLC, mMTC) with much higher performance. To meet requirements such as peak data rate of 1Tbps, user experienced data rate of 1Gbps, and reliability of $10^{-7}$, a suite of new technologies is needed. Reconfigurable intelligent surface (RIS) is one of the technologies envisioned to play a key role in 6G [3]-[6]. A RIS is a large programmable surface with elements that can be tuned to different reflection properties [3]. It can be placed in a wireless propagation environment as an artificial reflector to boost the signal from the transmitter to the receiver. This can improve the performance at the receiver as well as reduce the interference to other users. RIS-assisted wireless communication has gained a lot of attention recently. RIS-assisted SISO/MISO systems have been studied in [7][8][9], and MIMO systems have been studied in [10] [11]. RIS-assisted multi-user or multi-cell systems are reported in [11][12]. RIS with hybrid beamforming was studied in [13]. Combination of RIS with multiple access scheme has also been studied [14]. State-of-the-art of researches in this area can be found in some very comprehensive survey papers [15][16][17].

In this paper we focus on very high frequency range such as mmW and THz. THz is bound to be a key part of 6G spectrum due to large amount of available bandwidth [18]. At mmW or THz, diffraction is giving way to reflection and the channel is mostly comprised of discrete paths including the LOS path and some specular reflection paths [19][20]. Due to large propagation loss, large antenna array is required for high directional gain at the transmitter (TX) and the receiver (RX). An antenna array may have thousands of elements or more. As the size increases, the spatial/angular resolution of the antenna array improves [21]. Large antenna array combined with discrete channel makes the MIMO channel sparse in some bases, and the sparsity can be used to simplify the analysis.

The contribution of this paper is as follows. We explore the high spatial resolution of large antenna array and the discrete structure of multi-path channel. By decomposing the channel into a set of orthogonal spatial modes, we can maximize the channel capacity by multiplexing using these modes. The RIS surface can be used as a whole to support rank-1 reflection with maximal SNR, or can be partitioned into multiple subarrays for high rank transmission. The optimization is performed in the singular space of the channel matrix and on the TX power and the RIS area allocated to each beam (instead of on each RIS element as in many other papers), greatly reducing the complexity. We develop efficient algorithms to optimize the transmission rank, the set of beams used and their transmission powers, the TX & RX beamforming vectors, and the configuration of the RIS surface. These algorithms are verified with the MATLAB optimization toolbox. We also compared the traditional MIMO channel with the reflection channel and explained why it is harder to support high rank transmission in the latter.

The organization of the paper is as follows. We start with the multi-path sparse channel model in Section 2, and analyze a direct TX-RX link in Section 3. We then move on to the reflection channel on a RIS surface in Section 4, starting with rank-1 transmission /reflection and ending with multi-rank transmission/reflection. We show the detailed operations of the algorithms with examples. The results are verified with the MATLAB Optimization Toolbox. Section 5 concludes the paper.

A few notes on the notation: a scaler is represented in italic lower case ($a$), a vector in bold italic lower case ($\boldsymbol{a}$), a matrix in bold italic upper case ($\boldsymbol{A}$). $\boldsymbol{A}(i,:)$ or $\boldsymbol{a}_i^r$, and $\boldsymbol{A}(:,j)$ or $\boldsymbol{a}_j^c$ respectively represent the $i$-th row and $j$-th column of matrix $\boldsymbol{A}$. Algorithms are provided in MATLAB style pseudocodes in Appendix B.

## 2. Preliminary: sparse multi-path channel model

In the mmW and THz range, the channel can be characterized by a set of discrete, LOS or specular reflection paths, each with a set of unique properties [19][20]. Because most RIS devices are polarization sensitive[4], we only consider linearly polarized EM wave. In this section, we first introduce the channel model developed by Sayeed [22] before applying it to the RIS reflection channel.

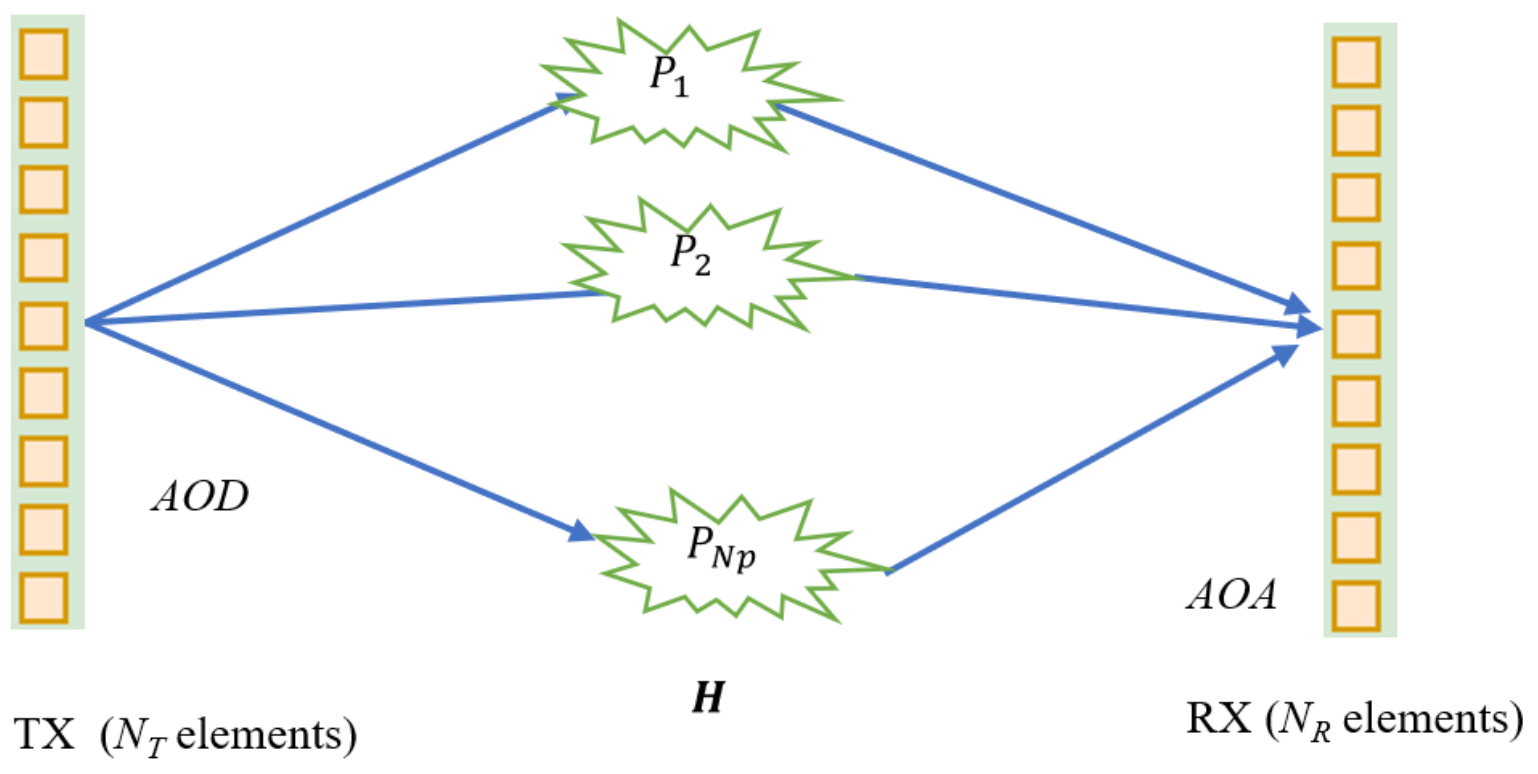

*Figure 1*. Discrete multi-path MIMO channel model.

Assume the channel between the TX and the RX consists of $N_p$ distinct paths. Ignoring the Doppler effect, path $n$ can be described by $(\beta_n, \theta_n^T, \theta_n^R, \tau_n)$, where $\beta_n$ is the complex channel coefficient, $\theta_n^T$ the normalized angle of departure (AOD) at the TX, $\theta_n^R$ the normalized angle of arrival (AOA) at the RX, $\tau_n$ the relative delay. We assume any two paths do not share the same value of $\theta_n^R$. The effective channel between the TX and the RX is $\boldsymbol{H}(f) = \sum_{n=1}^{N_p} \beta_n e^{-j2\pi\tau_n f} \boldsymbol{a}_R(\theta_n^R) \boldsymbol{a}_T{}^H(\theta_n^T)$. The $N_T \times 1$ vector $\boldsymbol{a}_T(\theta^T)$ is the spatial/angular response vector of the TX antenna array in direction $\theta^T$, and $N_R \times 1$ vector $\boldsymbol{a}_R(\theta^R)$ is the spatial/angular response vector of the RX antenna array in direction $\theta^R$. Detailed $\boldsymbol{a}_T(\theta^T)$, $\boldsymbol{a}_R(\theta^R)$ depend on the antenna configurations. For one dimensional uniform linear array (ULA), $\boldsymbol{a}_T(\theta^T) = \frac{1}{\sqrt{N_T}}\left[1, e^{-j\theta^T}, \ldots, e^{-j(N_T-1)\theta^T}\right]^T$, $\boldsymbol{a}_R(\theta^R) = \frac{1}{\sqrt{N_R}}\left[1, e^{-j\theta^R}, \ldots, e^{-j(N_R-1)\theta^R}\right]^T$, $\theta^T$ and $\theta^R$ are related to the physical AOD ($\phi^T$) and AOA ($\phi^R$) by $\theta^T = \frac{2\pi d_T \sin\phi^T}{\lambda}, \theta^R = \frac{2\pi d_R \sin\phi^R}{\lambda}$. Respectively $N_T$, $N_R$, $d_T$, $d_R$ are the number of antenna elements at the TX and RX, and their interspaces. Assume $N_T$, $N_R$ are large enough (we will define "large" soon), and $d_T$, $d_R$ are small enough ($\leq \lambda/2$). For subband $i$, $\boldsymbol{H}(f_i) = \sum_{n=1}^{N_p} \beta_n^i \, \boldsymbol{a}_R(\theta_n^R)\boldsymbol{a}_T{}^H(\theta_n^T)$, $\beta_n^i = \beta_n e^{-j2\pi\tau_n f_i}$. With OFDM, we can concentrate on a subband without loss of generality and drop the index $i$. The gives the generic channel $\boldsymbol{H} = \sum_{n=1}^{N_p} \beta_n \, \boldsymbol{a}_R(\theta_n^R)\boldsymbol{a}_T{}^H(\theta_n^T)$. A virtual channel model is a quantized approximation by uniformly sampling in the AOD/AOA space at their respective sampling rates $\Delta\theta^T = \frac{2\pi}{N_T}$, $\Delta\theta^R = \frac{2\pi}{N_R}$. In the virtual channel, $\boldsymbol{H}$ can be approximated as $\boldsymbol{H} \approx \sum_{i=0}^{N_R-1} \sum_{k=0}^{N_T-1} H_v(i,k)\boldsymbol{a}_R\left(\frac{2\pi i}{N_R}\right)\boldsymbol{a}_T{}^H\left(\frac{2\pi k}{N_T}\right) = \boldsymbol{A}_R \, \boldsymbol{H}_v \boldsymbol{A}_T^H$, $H_v(i,k) = \boldsymbol{a}_R^H\left(\frac{2\pi i}{N_R}\right)\boldsymbol{H}\boldsymbol{a}_T\left(\frac{2\pi k}{N_T}\right)$ is the path coefficient from TX direction $k$ to RX direction $i$. For the sake of simplicity, we use 1D ULA as an example. The results can be generalized to 2D uniform planar array by using 2D DFT and rearranging the matrices/vectors properly. With 1D ULA, $\boldsymbol{A}_T = \boldsymbol{DFT}_{N_T}$, $\boldsymbol{A}_R = \boldsymbol{DFT}_{N_R}$, where $\boldsymbol{DFT}_N$ is the DFT matrix of size $N$**.** The $N_P$ physical paths are partitioned into $N_R N_T$ bins in the AOD/AOA domain. Define the subset of paths in the angular domain $S_{i,k}^{RT} \triangleq S_i^R \cap S_k^T$ : $S_i^R \triangleq \left\{n\colon \theta_n^R \in \left(\frac{2\pi i}{N_R} - \frac{\pi}{N_R}, \frac{2\pi i}{N_R} + \frac{\pi}{N_R}\right]\right\}$, $S_k^T \triangleq \left\{n\colon \theta_n^T \in \left(\frac{2\pi k}{N_T} - \frac{\pi}{N_T}, \frac{2\pi k}{N_T} + \frac{\pi}{N_T}\right]\right\}$. As $N_T$, $N_R$ increase, $H_v(i,k) = \sum_{n=1}^{N_p} \beta_n f_{N_R}\left(\frac{2\pi i}{N_R} - \theta_n^R\right) f_{N_T}\left(\frac{2\pi k}{N_T} - \theta_k^T\right) \approx \sum_{n \in S_{i,k}^{RT}} \beta_n$, where $f_N(\theta)$ is the Dirichlet kernel $\frac{1}{N}\sum_{n=0}^{N-1} e^{-jn\theta}$. The approximation becomes true as $N_T, N_R \rightarrow \infty$. When the TX/RX arrays are large enough, the AOD/AOA bin becomes small enough for each $H_v(i,k)$ to contain at most one non-zero $\beta_n$, i.e. different paths are fully resolved. The assumption that not two paths have the same AOA implies that there is at most one non-zero element in each row of $\boldsymbol{H}_v$. Consequently the matrix $\boldsymbol{H}_v$ can be called row-sparse. A beam sent in AOD $\theta_n^T$ from the TX may be received in more than one AOAs at the RX, but each AOA corresponds to at most one AOD. We have the following theorem:

**Theorem 1**: For a row-sparse matrix $\boldsymbol{H}_v$, there exists a SVD decomposition $\boldsymbol{H}_v = \boldsymbol{USV^H}$ such that $\boldsymbol{V}$ is the identity matrix. (The proof is in Appendix A.)

This implies different TX beams are orthogonal, $\boldsymbol{H}_v = \boldsymbol{US}$. Define the set $S_T \triangleq \{\forall\ 0 \le k < N_T, |\boldsymbol{H}_v(:,k)| > 0\}$. These are TX beams or AODs with non-zero gains.

## 3. Direct TX to RX channel

When the TX transmits signal $\boldsymbol{x}$ using precoding matrix $\boldsymbol{P}$, the received signal is $\boldsymbol{y} = \boldsymbol{HPx} + \boldsymbol{n}_R$. Assume $\boldsymbol{x}$ is independent and normalized, and the receiver noise $\boldsymbol{n_R}$ has power $\sigma^2$. Using RX spatial filter $\boldsymbol{W}$, $\hat{\boldsymbol{y}} = \boldsymbol{Wy} = \boldsymbol{WHPx} + \boldsymbol{Wn}_R = \boldsymbol{WA}_R\boldsymbol{USA}_T^H\boldsymbol{Px} + \boldsymbol{Wn}_R = \boldsymbol{W_1USP_1x} + \boldsymbol{Wn}_R$, where $\boldsymbol{W_1} = \boldsymbol{WA}_R, \boldsymbol{P_1} = \boldsymbol{A}_T^H\boldsymbol{P}$. When $\boldsymbol{P_1}$ is diagonal and $\boldsymbol{W_1} = \boldsymbol{U^H}$, $\boldsymbol{W_1USP_1} = \boldsymbol{SP_1}$ is diagonal. It follows $\hat{\boldsymbol{y}} = \boldsymbol{SP_1x} + \boldsymbol{U^HA}_R^H\boldsymbol{n}_R = \boldsymbol{SP_1x} + \tilde{\boldsymbol{n}}_R$. $\boldsymbol{W} = \boldsymbol{U}^H\boldsymbol{A}_R^H$, $\tilde{\boldsymbol{n}}_{\boldsymbol{R}} = \boldsymbol{Wn}_R = \boldsymbol{U}^H\boldsymbol{A}_R^H\boldsymbol{n}_R$, $E(\tilde{\boldsymbol{n}}_R\tilde{\boldsymbol{n}}_R^H) = \sigma^2\boldsymbol{I}_{N_R}$. Precoder $\boldsymbol{P} = \boldsymbol{A}_T\boldsymbol{P_1}$ allocates TX power to the data streams and sends them in the AODs with non-zero gains. $\boldsymbol{P_1} = diag\left(\left[\sqrt{sp_0}, \sqrt{sp_1}, \dots \sqrt{sp_{N_T-1}}\right]\right)$ is power scaling vector, $sp_j = 0$ for $j \notin S_T$. The data stream transmitted in direction $k \in S_T$ is given TX power $sp_k$. The receiver estimate is $\hat{\boldsymbol{x}} = \boldsymbol{P_1^H S}^H(\boldsymbol{SP_1P_1^HS}^H + \sigma^2\boldsymbol{I})^{-1}\hat{\boldsymbol{y}}$. Because $\boldsymbol{H_v}$ is row-sparse, the RX sees at most one TX beam in any AOA, while the signal sent in an AOD may be received in more than one AOAs. The receiver can take the signals received in these AOAs and perform maximal ratio combining. The total received power of the signal sent in TX direction *k* is $sp_k|h_k^c|^2$, where $h_k^c$ is the *k*-th column of matrix $\boldsymbol{H_v}$, $|h_k^c| = s_k$ is the corresponding singular value.

We have $|S_T|$ parallel channels, where the signal sent in direction $k$ has $SNR_k = \frac{sp_k {s_k}^2}{\sigma^2}$. When the channel matrix $\boldsymbol{H_v}$ is known to the transmitter, it can determine a subset of $J$ beams from $S_T$ and their powers using the standard water filling algorithm, subject to the total power constraint $\sum_{j=1}^{J} sp_j = N_T p_{max}$, where $p_{max}$ is the maximal power of each transmission antenna. The transmitted signal is

$$\boldsymbol{x_p} = \boldsymbol{Px} = \boldsymbol{DFT}_{N_T}\boldsymbol{P_1 x} = \sum_{j=0}^{N_T-1} \boldsymbol{v_{N_T}^j}\sqrt{sp_j}x_j, \tag{1}$$

where $\boldsymbol{v_{N_T}^j}$ is the $j$-th column of the $\boldsymbol{DFT}_{N_T}$ matrix. The transmission power is

$$\begin{aligned} E(\boldsymbol{x_p x_p^H}) &= \left( \boldsymbol{DFT}_{N_T} E\left( \begin{bmatrix} \sqrt{sp_0}x_0 \\ \sqrt{sp_1}x_1 \\ \cdots \\ \sqrt{sp_{N_T-1}}x_{N_T-1} \end{bmatrix} \begin{bmatrix} \sqrt{sp_0}x_0 \\ \sqrt{sp_1}x_1 \\ \cdots \\ \sqrt{sp_{N_T-1}}x_{N_T-1} \end{bmatrix}^H \right) \boldsymbol{DFT}_{N_T}^H \right) \\ &= \left(\frac{1}{N_T}\sum_{j=0}^{N_T-1} sp_j\right) \boldsymbol{I_{N_T}} = p_{max}\boldsymbol{I_{N_T}}. \end{aligned} \tag{2}$$

Because the TX power is uniform across the antennas to fully utilize their PA powers, both the ULA antenna array and the precoder are optimal.

## 4. RIS reflection channel

The reflection channel from the TX to the RX through the RIS can be modeled as the product of three parts $\boldsymbol{H_2 H_R H_1}$. $\boldsymbol{H_1}$ is the channel from the TX to the RIS, $\boldsymbol{H_2}$ from the RIS to the RX, and $\boldsymbol{H_R}$ is the reflection of the RIS surface (Figure 2). Assume $\boldsymbol{H_1}$ and $\boldsymbol{H_2}$ are row-sparse. To simplify the analysis, we model each RIS element as an ideal phase shifter with unit omni-directional gain whose individual phase can be tuned arbitrarily. $\boldsymbol{H_D}$

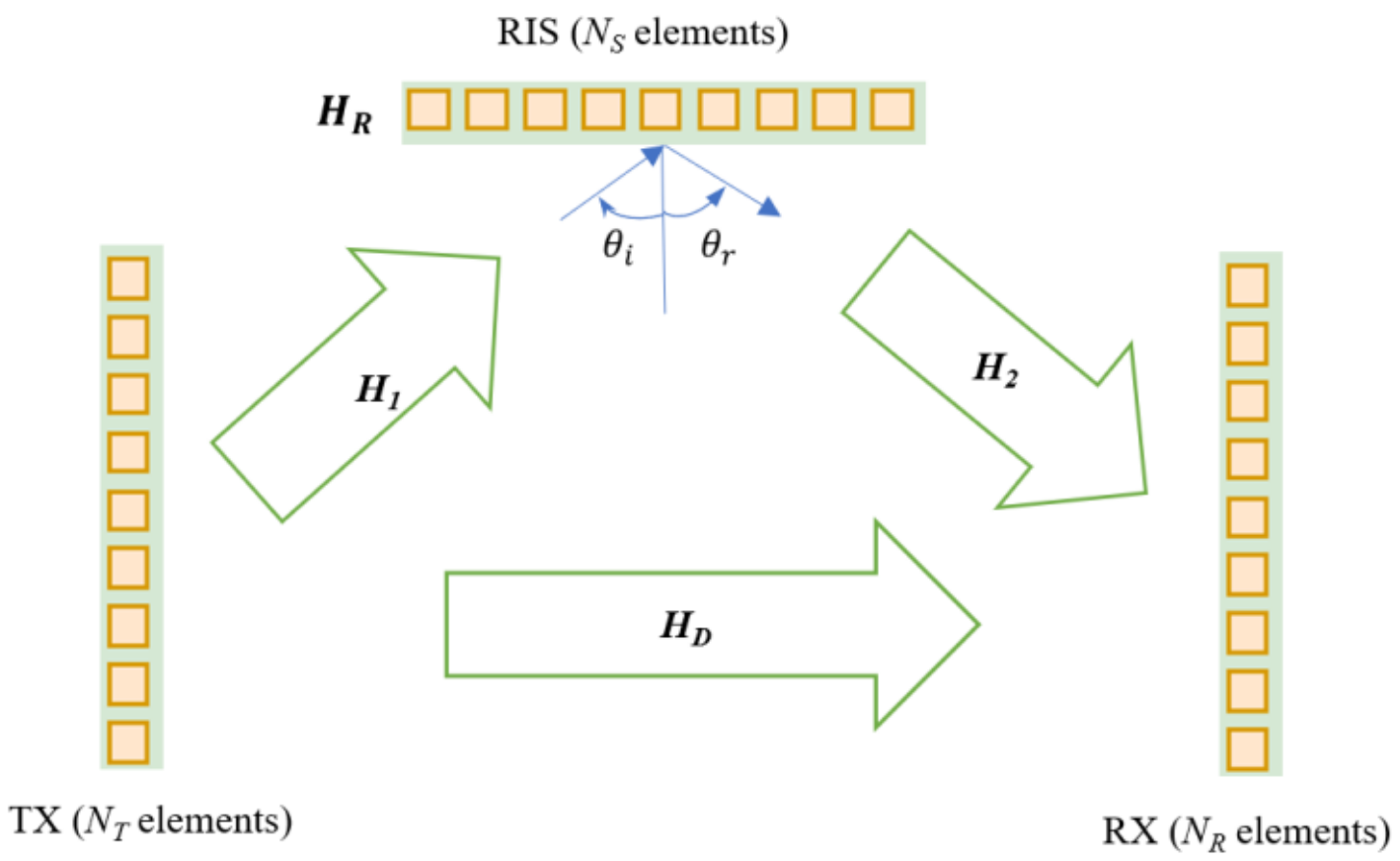

*Figure 2.* Channel model with a RIS surface.

is the direct channel from the TX to the RX. First we assume $\boldsymbol{H_D} = 0$. We will address $\boldsymbol{H_D} \neq 0$ later.

### 4.1. **Rank-1 reflection**

Assume the RIS surface consists of a 1D uniform array of $N_S$ elements numbered sequentially. The reflection coefficients of the $N_S$ elements are functions of phase vector $\boldsymbol{\phi_R} = \left[0, \phi_R^1, \ldots, \phi_R^{N_S-1}\right]$. The encoding vector for the RIS surface is $\boldsymbol{v}(\boldsymbol{\phi_R}) = e^{j\boldsymbol{\phi_R}} = \left[1, e^{j\phi_R^1}, e^{j\phi_R^2}, \ldots, e^{j\phi_R^{N_S-1}}\right]$. The RIS reflection is $\boldsymbol{H_R}(\boldsymbol{\phi_R}) = diag\left(\boldsymbol{v}(\boldsymbol{\phi_R})\right)$. The channel $\boldsymbol{H} = \boldsymbol{H_2 H_R H_1}$ becomes

$$\begin{aligned} \boldsymbol{H}(\boldsymbol{\phi_R}) = \boldsymbol{H_2 H_R}(\boldsymbol{\phi_R})\boldsymbol{H_1} &= \boldsymbol{DFT_{N_R} H_{v2} DFT^H_{N_S} H_R}(\boldsymbol{\phi_R})\boldsymbol{DFT_{N_S} H_{v1} DFT^H_{N_T}} \\ &= \boldsymbol{DFT_{N_R} H_{v2} {H_R}^+}(\boldsymbol{\phi_R})\boldsymbol{H_{v1} DFT^H_{N_T}}, \end{aligned} \quad (3)$$

where $\boldsymbol{H_R}^+(\boldsymbol{\phi_R}) = \boldsymbol{DFT^H_{N_S} H_R}(\boldsymbol{\phi_R})\boldsymbol{DFT_{N_S}}$. Let's take a close look at $\boldsymbol{H_{v2} H_R}^+(\boldsymbol{\phi_R})\boldsymbol{H_{v1}}$. When $\boldsymbol{\phi_R} = \boldsymbol{0}$ and $\boldsymbol{H_R}^+(\boldsymbol{\phi_R}) = \boldsymbol{I}$, the RIS surface performs ideal mirror reflection, taking an incoming beam from the TX with incident angle $\theta_i$ and bouncing it off with reflection angle $\theta_r = \theta_i$. With thousands of elements or more in the TX, RIS and RX, and only a few paths between them, $\boldsymbol{H_{v1}}$ and $\boldsymbol{H_{v2}}$ are very sparse. There is a good chance that $\boldsymbol{H_{v2}H_{v1}}$ produces an all-zero matrix, i.e. all incident beams from the TX in $\boldsymbol{H_{v1}}$ are reflected towards null directions of $\boldsymbol{H_{v2}}$. To prevent this from happening, $\boldsymbol{H_R}$ needs to manipulate the reflection direction of an incident beam and send it towards a non-null direction of $\boldsymbol{H_{v2}}$. This requires $\boldsymbol{H_R}^{+}(\boldsymbol{\phi_R})$ to take the form of a cyclic shift matrix $\boldsymbol{C}_{N_S}(N_C)$ for some integer $0 \leq N_C < N_S$, where $\boldsymbol{C}_N(k)$ is the matrix obtained by cyclic shifting the identity matrix of size $N$ up by $k$ rows. The makes the RIS reflection

$$\boldsymbol{H_R}(\phi_R) = \boldsymbol{DFT_{N_S}}\boldsymbol{C}_{N_S}(N_C)\boldsymbol{DFT_{N_S}^{H}} = diag\left(\left[1, e^{j\phi_c}, e^{j2\phi_c}, \dots, e^{j(N_S-1)\phi_c}\right]\right) \quad (4)$$

$\boldsymbol{v}(\boldsymbol{\phi_R})$ is a DFT vector, $\phi_c = -2\pi N_C/N_S$. The channel $\boldsymbol{H}$ becomes a function of $N_C$:

$$\boldsymbol{H}(N_C) = \boldsymbol{DFT_{N_R}}\boldsymbol{H_{v2}}\boldsymbol{C}_{N_S}(N_C)\boldsymbol{H_{v1}}\boldsymbol{DFT_{N_T}^{H}}. \quad (5)$$

$\boldsymbol{C}_{N_S}(N_C)$ first cyclic shifts $\boldsymbol{H_{v1}}$ up by $N_C$ rows before multiplying it with $\boldsymbol{H_{v2}}$. The RIS deflects the incident beams by $2\pi N_C/N_S$ to match them with the beams towards the RX. Because $\boldsymbol{H_{v1}}$ and $\boldsymbol{H_{v2}}$ are both sparse, we are unlikely to find a value of $N_C$ to match multiple pairs of beams simultaneously. The optimal encoding strategy for the RIS is to find the strongest incident beam $i_1^*$ in $\boldsymbol{H_{v1}}$, $(i_1^*, k_1^*) = argmax_{(i,k)} |\boldsymbol{H_{v1}}(i,k)|$, and match it with the strongest beam $k_2^*$ in $\boldsymbol{H_{v2}}$, $k_2^* = argmax_k |\boldsymbol{H_{v2}}(:,k)|$. Because $\boldsymbol{H_{v2}}$ is row-sparse, the signal sent by the TX in direction $k_1^*$ and reflected on the RIS may be received in more than one RX directions. To match $i_1^*$ to $k_2^*$, $N_c^* = k_2^* - i_1^*$ is used to determine the phases of the RIS elements. For RIS with discrete phase, the closest value at each element with the smallest quantization error can be used. $\boldsymbol{H}_{v2}(:,k_2^*)\boldsymbol{C}_{N_S}(k_2^* - i_1^*)\boldsymbol{H}_{v1}(i_1^*, k_1^*)$ is the resulting rank-1 channel, $g_R{}^* = |\boldsymbol{H}_{v2}(:,k)|^2 |\boldsymbol{H}_{v1}(i_1^*, k_1^*)|^2$ is the effective channel gain. $|\boldsymbol{H}_{v2}(:,k_2^*)| = s_{2,k_2^*}$ is the largest singular value of $\boldsymbol{H_{v2}}$. The choice of $(i_1^*, k_2^*)$ maximizes $|\boldsymbol{H}_{v1}(i_1^*, k_1^*)|$ and $|\boldsymbol{H}_{v2}(:, k^*)|$ respectively, so the gain ${g_R}^*$ is maximized. The reflection degenerates the channel into a rank-1 channel with the largest gain to maximize the SNR, even when $\boldsymbol{H_{v1}}$ and $\boldsymbol{H_{v2}}$ both have higher ranks. The analysis of 1D array can be extended to 2D by replacing 1D DFT with 2D DFT and with proper rearrangement of the matrices/vectors. Effectively two separate 1D reflections apply to the X and Y directions. The parameter $\phi_c$ is replaced with a pair $(\phi_c^x, \phi_c^y)$, representing the reflection in the X and Y directions respectively.

When $\boldsymbol{H_D} \neq \boldsymbol{0}$ , the channel $\boldsymbol{H} = \boldsymbol{H_2}\boldsymbol{H_R}(\boldsymbol{\phi_R})\boldsymbol{H_1} + \boldsymbol{H_D}$. Assume $\boldsymbol{H_1}$ and $\boldsymbol{H_D}$ do not share any common AODs, and $\boldsymbol{H_2}$ and $\boldsymbol{H_D}$ do not share any common AOAs. From the above analysis, the RIS surface reflects a single incident beam towards the RX, reducing the reflection channel $\boldsymbol{H_2}\boldsymbol{H_R}(\boldsymbol{\phi_R})\boldsymbol{H_1}$ to a rank-1 channel. This reflected beam is orthogonal to all the beams of $\boldsymbol{H_D}$. The rank of $\boldsymbol{H}$ is higher than $\boldsymbol{H_D}$ by one, with all columns mutually orthogonal. The standard water filling algorithm can be applied to all the beams of $\boldsymbol{H}$, including the direct beams and the reflected beam to maximize the capacity.

### 4.2. Higher rank reflection

In this section we try to increase the rank of the reflection to improve the capacity. We start with the case of $\boldsymbol{H_D} = \boldsymbol{0}$. Section 4.1 demonstrated that a single beam can be reflected when a DFT vector is applied to the entire RIS surface. To support higher rank reflection, we partition the RIS surface into $J$ disjoint subarrays. Each subarray consists of a subset of adjacent elements from the original RIS surface and reflects a single beam from the TX to the RX. We assume the RIS array is large enough so after splitting each subarray still has enough elements and spatial resolution to resolve all the AOAs and AODs. The channel sparsity is preserved at each subarray. Define the relative size of subarray $j$ by $r_j$, the $j$-th subarray has $r_j N_S$ elements[1]. We model $r_j$ as a real number $0 \le r_j \le 1$, $\sum_{j=1}^{J} r_j = 1$. Each subarray functions as a small independent rank-1 reflector, sending an incident beam from the TX towards the RX. Following Sec 4.1, the elements of a subarray are encoded as a DFT vector with parameter $\phi_c^j$: $\boldsymbol{v_j}(\phi_c^j) = \left[1, e^{j\phi_c^j}, e^{j2\phi_c^j}, \dots, e^{j(r_j N_S - 1)\phi_c^j}\right]$. The entire RIS surface reflection $\boldsymbol{H_R}(\boldsymbol{r}, \boldsymbol{\phi_C}) = diag([\boldsymbol{v_1}(\phi_c^1), \dots, \boldsymbol{v_J}(\phi_c^J)])$ is a function of the subarray size $\boldsymbol{r} = [r_1, \dots, r_J]$ and the phase vector $\boldsymbol{\phi_C} = [\phi_c^1, \dots, \phi_c^J]$.

We now develop an algorithm to maximize the total capacity with these beam pairs. We optimize the transmission rank, the set of TX beams and their transmission powers, and the configurations of the RIS subarrays including their relative sizes ($\boldsymbol{r}$) and phases ($\boldsymbol{\phi_C}$). From Section 3, the TX can generate $J$ beams simultaneously, transmitting the $j$-th beam in the direction of $k_1^j$ using power $p_j$. On the RIS surface, the $j$-th subarray reflects the $j$-th beam incoming from direction $i_1^j$ and sends it towards direction $k_2^j$ with phase $N_c^j = k_2^j - i_1^j$. Here consider the beam directions $i_1^j$, $k_2^j$ and $N_c^j$ normalized with respect to the subarray size. The SNR of the $j$-th beam pair at the RX is a function of gain $g_j$, subarray size $r_j$ and TX power $p_j$. The array response function is linear with respect to the size of the array [21], so the SNR of the signal reflected on subarray $i$ is proportional to ${r_j}^2$:

$$SNR_j(r_j, p_j) = \frac{g_j {r_j}^2 p_j}{\sigma^2} = \frac{g_j {r_j}^2 q_j P_{total}}{\sigma^2} = {r_j}^2 q_j SNR_j^o \quad (6)$$

where $P_{total}$ is the total transmission power density. The ratio $0 \le q_j \le 1$ is the ratio of the TX power used for beam $j$, and $\sum_{j=1}^{J} q_j = 1$. $SNR_j^o = \frac{g_j P_{total}}{\sigma^2}$ is the normalized SNR of the $j$-th beam pair when all the TX power and the entire RIS surface are assigned to it. For bandwidth $W$, the capacity of a beam pair is given by the Shannon capacity. The goal is to find the $(\boldsymbol{r}^*, \boldsymbol{q}^*)$ to maximize the total capacity of $J$ beam pairs **(P1)**:

$$\max_{\boldsymbol{r},\boldsymbol{q}} C(\boldsymbol{r},\boldsymbol{q}) = \max_{\boldsymbol{r},\boldsymbol{q}} \sum_{j=1}^{J} W \log_2\left(1 + {r_j}^2 q_j SNR_j^o\right) \tag{7}$$

$$s.t. \quad \sum_{j=1}^{J} q_j = \sum_{j=1}^{J} r_j = 1\,,\ q_j \geq 0\ ,\ r_j \geq 0. \tag{8}$$

[1] For 2D array, subarray $j$ span $r_j^x N_S^x$ and $r_j^y N_S^y$ elements in the X and Y directions, and its relative size is $r_j = r_j^x r_j^y$.

Unlike the classic problem of power allocation in parallel AWGN channels, there is no closed form solution. To gain some insight, let's look at a slightly different problem, where we introduce a new variable $\boldsymbol{p}$ and replace ${r_j}^2 q_j$ with $r_j p_j q_j$ (**P2)**:

$$\max_{\boldsymbol{r},\boldsymbol{p},\boldsymbol{q}} C(\boldsymbol{r},\boldsymbol{p},\boldsymbol{q}) = \max_{\boldsymbol{r},\boldsymbol{p},\boldsymbol{q}} \sum_{j=1}^{J} W \log_2\left(1 + r_j p_j q_j SNR_j^o\right) \tag{9}$$

$$s.t. \quad \sum_{j=1}^{J} r_j = \sum_{j=1}^{J} p_j = \sum_{j=1}^{J} q_j = 1\,,\ r_j \geq 0,\ p_j \geq 0,\ q_j \geq 0. \tag{10}$$

Using Lagrangian multiplier, we get

$$\frac{p_j q_j SNR_j^o}{1 + r_j p_j q_j SNR_j^o} = \lambda_1, \frac{r_j p_j SNR_j^o}{1 + r_j p_j q_j SNR_j^o} = \lambda_2, \frac{r_j q_j SNR_j^o}{1 + r_j p_j q_j SNR_j^o} = \lambda_3. \tag{11}$$

Dividing these three equations by each other gives $\frac{q_j}{r_j} = \frac{\lambda_1}{\lambda_2}$ and $\frac{p_j}{r_j} = \frac{\lambda_1}{\lambda_3}$. This implies $\boldsymbol{r}^*$, $\boldsymbol{p}^*$ and $\boldsymbol{q}^*$ are proportional to each other. Given they are each normalized, it follows that $\boldsymbol{r}^* = \boldsymbol{p}^* = \boldsymbol{q}^*$. Consequently the solution of **P2** is identical to **P1**. We can take advantage of $\boldsymbol{r}^* = \boldsymbol{p}^* = \boldsymbol{q}^*$ and further simply **P2** to **P3:**

$$\max_{\boldsymbol{r}} C(\boldsymbol{r}) = \max_{\boldsymbol{r}} \sum_{j=1}^{J} W \log_2\left(1 + r_j^3 SNR_j^o\right), \tag{12}$$

$$s.t. \quad \sum_{j=1}^{J} r_j = 1, r_j \geq 0. \tag{13}$$

We start by plotting the total capacity for 2 beam pairs with different SNRs (Figure 3). It turns out the capacity function is neither convex nor concave. At low SNR (bottom), the capacity curve appears to be convex, with a maximum at each end. As SNR increases, a local maximum gradually grows in the middle. At high SNR (top), a maximum in the center is accompanied by a local minimum on each side (but not at the end). The function at the boundary is still convex. This observation is repeated when we plot the capacity function for 3 beam pairs (Figure 4 and Figure 5). The (local or global) maximum near the center is surrounded by $J$ minimums. There are other maximums on the boundary of the feasible region with $r_j = 0$ for some $j$. One of them is the true global maximum.

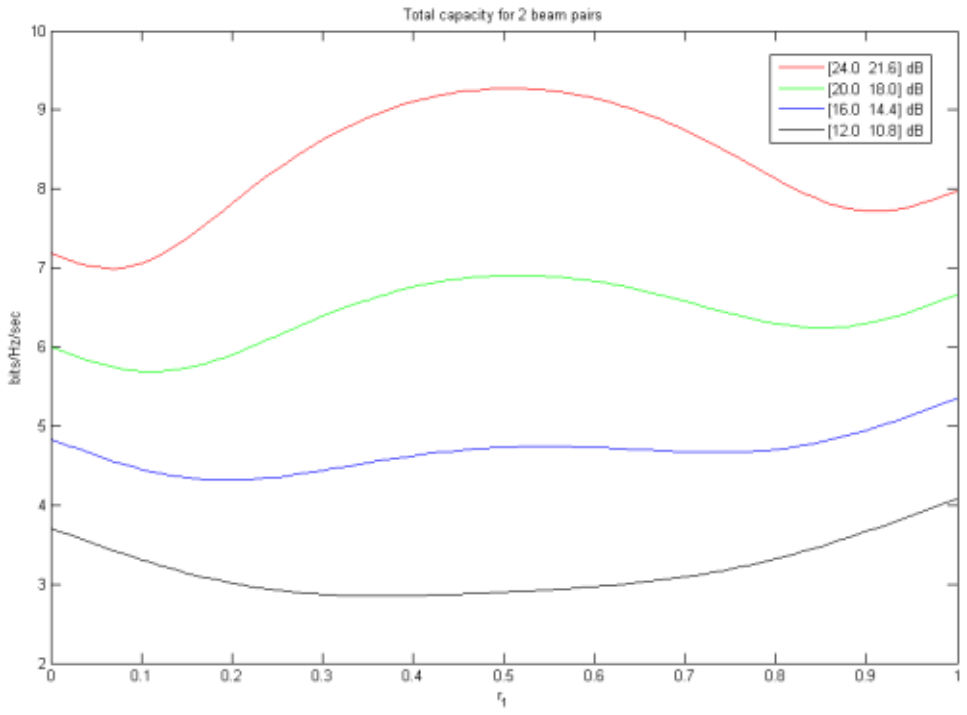

*Figure 3*. Capacity function of 2 beam pairs with different SNRs.

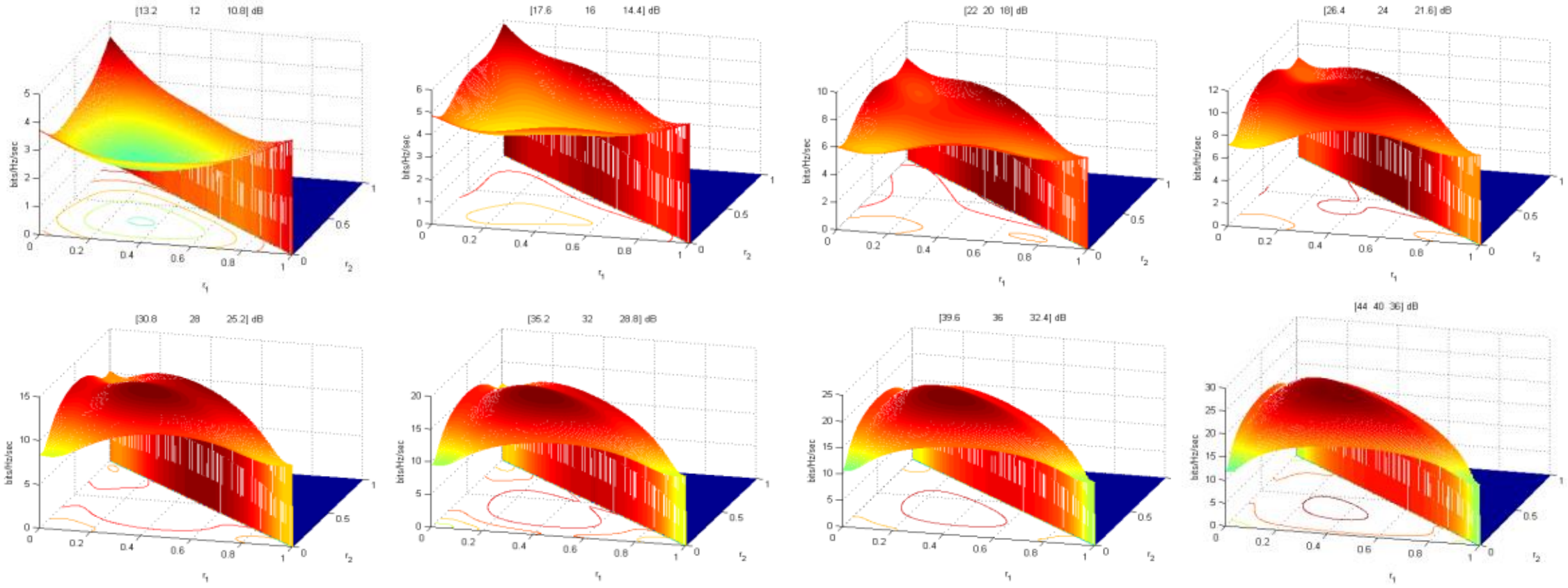

*Figure 4.* Capacity function of 3 beam pairs with different SNRs.

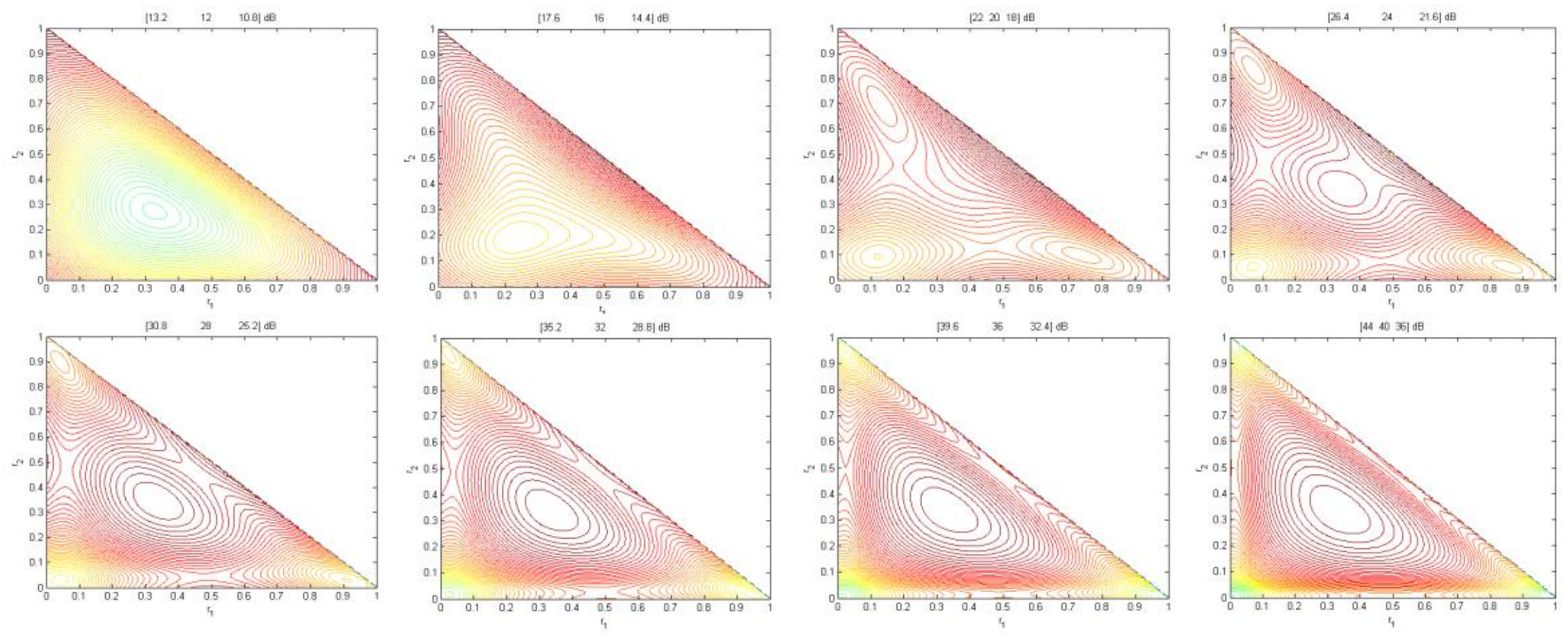

*Figure 5.* Contour plots of the capacity function corresponding to *Figure 4.*

We proceed to solve **P3**. If a maximum exists near the center, we can start from the center and find it using iterative search. We can improve the capacity by updating $\boldsymbol{r}$ using the water filling algorithm:

$$r_j^{(t)} = \left[\frac{1}{\lambda_2} - \frac{1}{{r_j^{(t-1)}}^2 SNR_j^o}\right]^+ . \tag{14}$$

This is implemented in *Opt_RIS_rank*() of Algorithm 1. Starting from the centroid of the feasible region, the algorithm updates $\boldsymbol{r}$ iteratively until convergence. Figure *6* shows the onvergence of *Opt_RIS_rank*() for four beam pairs with normalized $SNR^o = [22, 21, 20,$ 19] dB. The convergence criterium $\epsilon_{conv} = 10^{-4}$. The weakest beam pair ($SNR^o$=19 dB) is dropped after 2 iterations. The algorithm converges after 17 iterations.

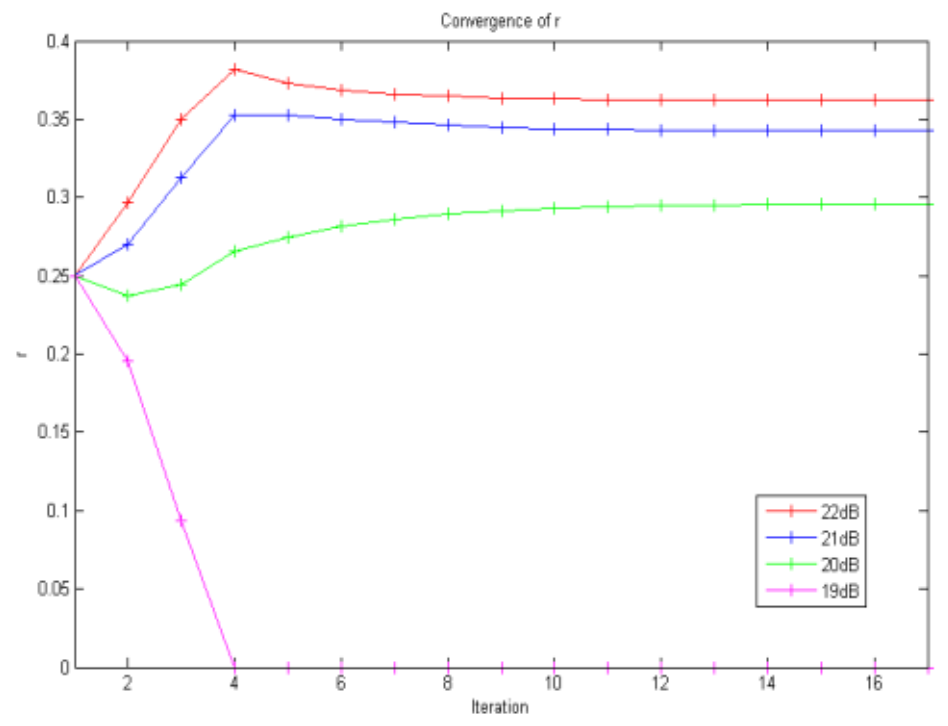

Figure 6. Convergence of *Opt_RIS_rank*($\boldsymbol{snr}$) for four beam pairs.

After *Opt_RIS_rank*() converges, if it lands on a maximum not on the boundary, it needs to be compared with the maximums on the boundary. However, if it ends in the boundary, there is no guarantee it has found the global maximum either. Because $r_j = 0$ for some $j$ on the boundary, finding the maximum on the boundary is equivalent to finding the maximum of a problem with reduced dimensions. We can include only a subset of beam pairs to see if the capacity increases. When choosing a subset, a stronger beam pair with higher SNR shall be chosen before a weaker beam pair. For rank *k*, we only need to include the strongest $k$ beam pairs instead of searching all the combinations with $k$ beam pairs. By inducting over $1 \leq k \leq J$ and comparing their capacities, the maximal capacity of the system can be found. The result is the *Joint_power_RIS_alloc_1*() algorithm. Table 1 shows *Joint_power_RIS_alloc_1*() with different number of beam pairs. Each time the strongest *k* beam pairs are included. Starting from the strongest beam pair and adding the next strongest beam pair one at a time, the capacity is computed for different $k$. It stops when no capacity improvement is achieved by adding more beam pairs. For illustration purposes we show $k = 1,2,3,4$ in the Table 1 and find $k$ = 2 gives the highest capacity. Adding the 3rd beam pair decreased the capacity, and the 4th beam pair did not get any power or reflection area. The algorithm stops after $k$ = 3 and determines the optimal transmission rank is 2. To validate the algorithm, we compare it with the interior point method of the MATLAB optimization toolbox (*fmincon*) with randomized starting points. The two methods produced identical results for a large number of randomly generated inputs. This verifies Algorithm 1 indeed produces the maximal capacity.

*Table 1*. Capacity and optimized $\boldsymbol{r}^*$ for different number of reflection beam pairs in *Joint_power_RIS_alloc_1*() of Algorithm 1.

| Number of beam pairs | SNR of selected beam pairs (dB) | Capacity (b/s/Hz) | $\boldsymbol{r}^*$ |
|---|---|---|---|
| 1 | [22] | 7.3173 | [1.0000] |
| 2 | [22,21] | 8.4444 | [0.5037, 0.4963] |
| 3 | [22,21,20] | 7.5295 | [0.3619, 0.3422, 0.2959] |
| 4 | [22,21,20,19] | 7.5295 | [0.3619, 0.3422, 0.2959, 0] |

We now show the beam pairing scheme is optimal at high SNR. Let $J^*$ be the number of beam pairs that are allocated non-zero resources. The capacity is approximately

$$C(\boldsymbol{r},\boldsymbol{q}) = \sum_{j=1}^{J^*} W \log_2\left(1 + {r_j}^2 q_j SNR_j^o\right) \approx \sum_{j=1}^{J^*} W \log_2\left({r_j}^2 q_j SNR_j^o\right)$$

$$= W\left(\sum_{j=1}^{J^*} \log_2\left(\frac{P_{tot}}{\sigma^2}\right) + \sum_{j=1}^{J^*} \log_2\left({r_j}^2 q_j\right) + \sum_{j=1}^{J^*} 2\log_2\left(\left|\boldsymbol{H}_{v2}(:,k_2^j)\right|\right) + \sum_{j=1}^{J^*} 2\log_2\left(\left|\boldsymbol{H}_{v1}(i_1^j,k_1^j)\right|\right)\right). \quad (15)$$

In the last step, the incident beam $i_1^j$ is decoupled from the outgoing beam $k_2^j$. As long as the $J^*$ strongest incident beams and the $J^*$ strongest outgoing beams are all included, it does not matter how they are paired. By pairing the *j*-th strongest incident beam with the *j*-th strongest outgoing beam, the beam pairing algorithm ensures the first $J^*$ beam pairs always have the $J^*$ strongest gains, for any value of $J^*$. The capacity is maximized for any transmission rank. Consequently, the beam pairing algorithm is optimal for large SNR.

Another observation at large SNR is that both the transmission power and the reflection area are distributed to the utilized beam pairs approximately evenly. The capacity

$$C(\boldsymbol{r},\boldsymbol{q}) = \sum_{j=1}^{J^*} W \log_2\left(1 + {r_j}^2 q_j SNR_j^o\right) \approx \sum_{j=1}^{J^*} W \log_2\left({r_j}^2 q_j SNR_j^o\right)$$
$$= W\left(\sum_{j=1}^{J^*} \log_2\left(SNR_j^o\right) + 2\log_2\left(\prod_{j=1}^{J^*} r_j\right) + \log_2\left(\prod_{j=1}^{J^*} q_j\right)\right). \quad (16)$$

Because of the constraint $\sum_{j=1}^{J^*} r_j = 1$, the product $\prod_{j=1}^{J^*} r_j$ is maximized when $r_j = \frac{1}{J^*}, 1 \leq j \leq J^*$. The same argument applies to $q_j$. Therefore, the capacity is maximized at the centroid $\boldsymbol{r^e} = \boldsymbol{q^e} = [1,1,\ldots,1]/J^*$. This is the reason we choose $\boldsymbol{r^e}$ as the starting point for *Opt_RIS_rank*(). This observation also leads us to simplify Algorithm 1 to Algorithm 2. In *Joint_power_RIS_alloc_2*(), we first compute the channel capacity under different ranks using $(\boldsymbol{r^e}, \boldsymbol{q^e})$ as an approximation to find the optimal rank ($rk^*$), then use *Opt_RIS_rank*() to find the optimal configuration $(\boldsymbol{r}^*, \boldsymbol{q}^*)$ under rank $rk^*$. In our simulations, *Joint_power_RIS_alloc_2*() produced the same results as *Joint_power_RIS_alloc_1*().

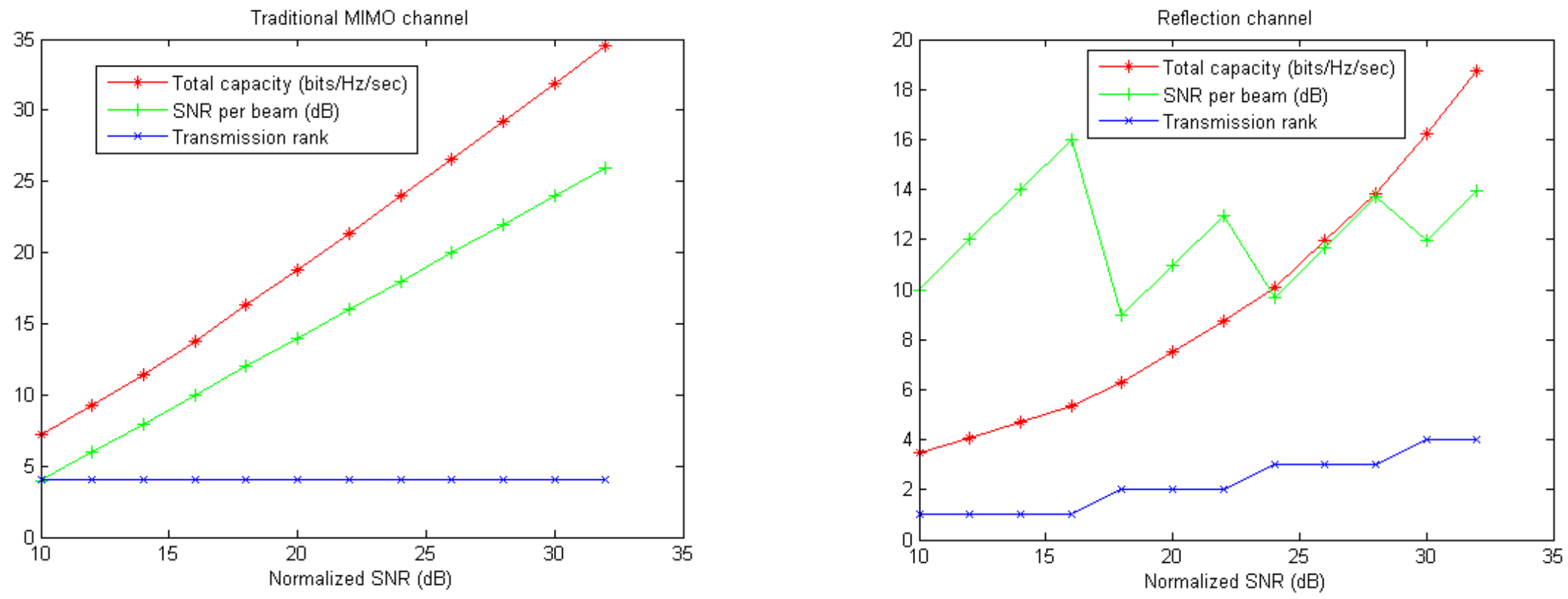

*Figure 7*. Transmission in the traditional MIMO channel (left) and the RIS reflection channel (right) with 4 beams or beam pairs of equal strength.

Figure 7 compares transmissions in traditional MIMO channel and RIS reflection channel. In both cases we assume there are 4 beams (beam pairs) with equal strength, and compute the optimal rank, the total capacity, and the effective SNR for each layer. Their stark difference deserves careful explanation. The traditional MIMO channel is limited by degree of freedom right from the beginning. Starting at 10 dB, it utilizes all the available degree of freedom and transmits at the maximal rank of 4. The capacity grows logarithmically with the SNR. In the RIS reflection channel, the transmission rank starts from 1 at 10 dB and slowly reaches 4 only at 30 dB. Most of the time it is limited by SNR. Because all the beams/beam pairs have equal strength, the resources are shared evenly across the transmitted layers ($k$). The power and RIS area allocated to each layer are $q_i = r_i = 1/k$. The effective SNR for each layer is $\propto 1/k$ for the MIMO channel and $\propto 1/k^3$ for the reflection channel. This makes the reflection channel much more sensitive to the transmission rank, and become SNR limited again each time the rank increments by 1. This makes it much harder to support high rank transmission in the RIS reflection channel than in the MIMO channel, particularly in the SNR range typically found in wireless networks. Our result cautions against using RIS for high rank reflection.

#### 4.2.1. Composite channel including direct and reflection channels

We now turn to $\boldsymbol{H} = \boldsymbol{H_D} + \boldsymbol{H_2 H_\phi H_1}, \boldsymbol{H_D} \neq \boldsymbol{0}$. Let $S_D$ be the set of direct beams in $\boldsymbol{H_1}$, and $S_R$ be the set of reflected beam pairs in $\boldsymbol{H_2 H_\phi H_1}$. Assume $\boldsymbol{H_1}$, $\boldsymbol{H_2}$ are row-sparse, $\boldsymbol{H_1}$ and $\boldsymbol{H_D}$ do not share any AODs, $\boldsymbol{H_2}$ and $\boldsymbol{H_D}$ do not share any AOAs. All the beams in $S_D \cup S_R$ are orthogonal to each other. We need to determine what direct and reflected beams to use and their TX powers, and the RIS configuration for the reflected beams. We first use the beam pairing algorithm to determine the reflected beam pairs $S_R$ and the RIS phase coefficients for the subarrays, then find the optimal transmission power allocation $\boldsymbol{q}^*$ for all the beams in $S_D \cup S_R$, and the reflection surface allocation $\boldsymbol{r}^*$ for the beam pairs in $S_R$. The problem to maximize the capacity is formulated as **P4:**

$$\max_{\boldsymbol{r},\boldsymbol{q}} C(\boldsymbol{r},\boldsymbol{q}) = \max_{\boldsymbol{r},\boldsymbol{q}} \left( \sum_{j \in S_R} W \log_2\left(1 + {r_j}^2 q_j SNR_j\right) + \sum_{j \in S_D} W \log_2\left(1 + q_j SNR_j\right) \right), \tag{17}$$

$$s.t. \quad \sum_{j \in S_D \cup S_R} q_j = \sum_{j \in S_R} r_j = 1\,, q_j \geq 0, r_j \geq 0. \tag{18}$$

Like **P3**, we begin by exploring the relationship between variable $\boldsymbol{q}$ and $\boldsymbol{r}$. By introducing a new variable $\boldsymbol{p}$ and converting **P4** to **P5**:

$$\max_{\boldsymbol{r},\boldsymbol{q}} C(\boldsymbol{r},\boldsymbol{q}) = \max_{\boldsymbol{r},\boldsymbol{q}} \left( \sum_{j \in S_R} W \log_2\left(1 + r_j p_j q_j SNR_j\right) + \sum_{j \in S_D} W \log_2\left(1 + q_j SNR_j\right) \right), \tag{19}$$

$$s.t. \quad \sum_{j \in S_D \cup S_R} q_j = \sum_{j \in S_R} r_j = \sum_{j \in S_R} p_j = 1\,, q_j \geq 0, r_j \geq 0, p_j \geq 0. \tag{20}$$

and using Lagrangian multiplier, we get

- For reflected beams $j \in S_R$,

$$\frac{r_j p_j SNR_j^o}{1 + r_j p_j q_j SNR_j^o} = \lambda_1, \frac{p_j q_j SNR_j^o}{1 + r_j p_j q_j SNR_j^o} = \lambda_2, \frac{r_j q_j SNR_j^o}{1 + r_j p_j q_j SNR_j^o} = \lambda_3. \tag{21}$$

Divide these equations by each other gives $\frac{r_j}{q_j} = \frac{\lambda_1}{\lambda_2}$ and $\frac{p_j}{q_j} = \frac{\lambda_1}{\lambda_3}, j \in S_R$. Because $\sum_{j \in S_R} r_j = \sum_{j \in S_R} p_j = 1$, it follows $r_j^* = p_j^* = \frac{q_j^*}{\sum_{i \in S_R} q_i^*},\ j \in S_R$.

- For direct beams $j \in S_D$,

$$\frac{SNR_j^o}{1+q_j SNR_j^o} = \lambda_1, j \in S_D. \tag{22}$$

Combine Eq(21)  and Eq(22), we get

$$q_j^* = \begin{cases} \left(\frac{1}{\lambda_1} - \frac{1}{SNR_j^o}\right)^+, & j \in S_D, \\ \left(\frac{1}{\lambda_1} - \frac{1}{r_j{}^* p_j{}^* SNR_j^o}\right)^+, & j \in S_R. \end{cases} \tag{23}$$

This becomes the foundation of the following iterative algorithm:

$$q_j^{(t)} = \begin{cases} \left(\frac{1}{\lambda_1} - \frac{1}{SNR_j^o}\right)^+, & j \in S_D, \\ \left(\frac{1}{\lambda_1} - \frac{1}{r_j{}^{(t-1)^2} SNR_j^o}\right)^+, & j \in S_R, \end{cases} \tag{24}$$

$$r_j^{(t)} = \frac{q_j^{(t)}}{\sum_{i \in S_R} q_i^{(t)}}, \ j \in S_R. \tag{25}$$

Eq(24) together with the constraint of Eq(20) can be solved by the standard water filling algorithm. Eq(25) just performs normalization. This is implemented in *Opt_Dir_RIS_rank*() in Algorithm 3. Figure 8 show *Opt_Dir_RIS_rank*() with 4 reflected beam pairs and 4 direct beams. The SNRs are [24, 22, 21, 20] dB for the reflected beams and [20, 19, 18, 17] dB for the direct beams. After 3 iterations, the two weakest reflected beams are dropped. All the direct beams and two remaining reflected beams are given TX power when the algorithm converges after 8 iterations. Different reflection ranks are compared in *Joint_power_Dir_RIS_alloc_1*(). Each time the strongest *k* reflected beams and all the direct beams are included to compute the capacity. *Joint_power_Dir_RIS_alloc_1*() with different number of reflected beam pairs is shown in *Table 2*. For k=3 or 4, the 3rd or 4th reflected beam pairs do not improve the capacity, and they are not given any TX power by *Opt_Dir_RIS_rank*(). We include $k$=3, 4 in *Table 2* only for illustration purpose. The algorithm would have terminated after $k$=2. The highest capacity is achieved at $k$=1, where only the strongest reflected beam and all four direct beams are used for transmission. The total capacity is 21.3817 bits/s/Hz with transmission rank of 5.

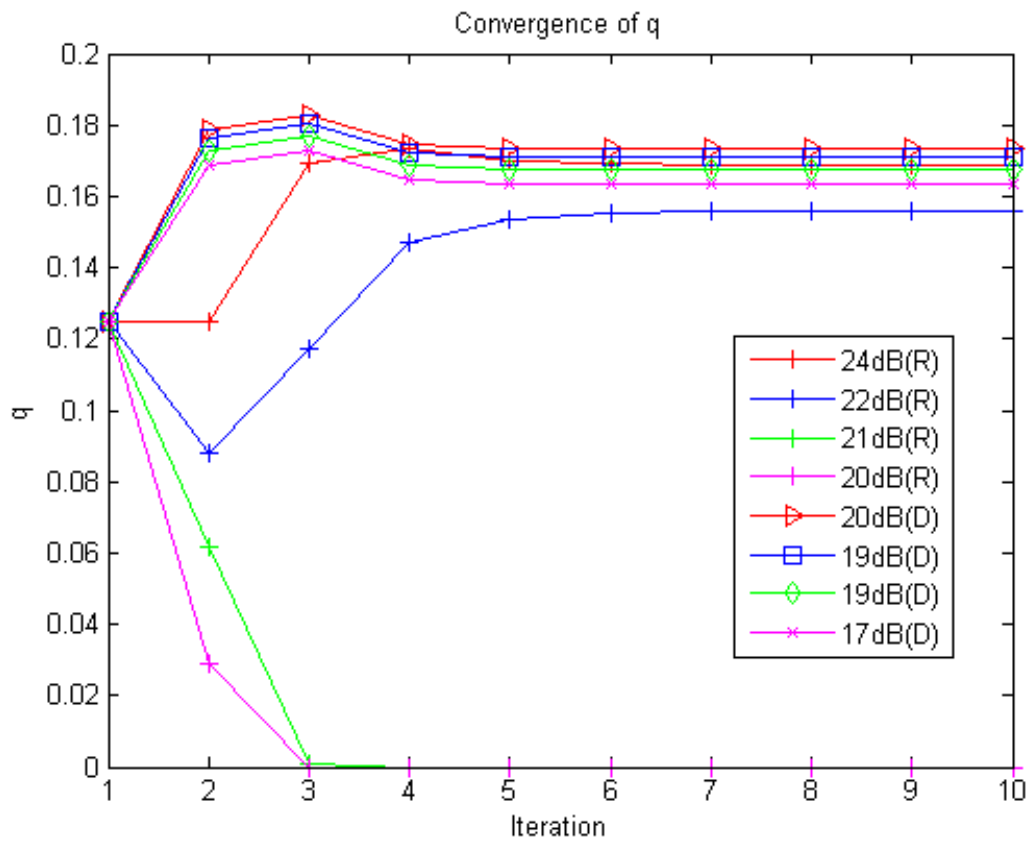

*Figure 8.* Convergence of the *Opt_Dir_RIS_rank*() of Algorithm 3. Reflected beams are labeled as (R) and direct beams are labeled as (D).

*Table 2.* Capacity and optimal reflection area ($\boldsymbol{r}^*$) and power allocation ($\boldsymbol{q}_R^*, \boldsymbol{q}_D^*$) with different number of reflected beams in *Joint_power_Dir_RIS_alloc_1*().

| Number of direct + reflected beams | SNR of direct & reflected beams (dB) | Capacity of reflected beams (bits/Hz/s) | Capacity of direct beams (bits/Hz/s) | Total capacity (bits/Hz/s) | Reflection area ($\boldsymbol{r}^*$) | TX power of reflected beams ($\boldsymbol{q}_R^*$) | TX power of direct beams ($\boldsymbol{q}_D^*$) |
|---|---|---|---|---|---|---|---|
| 4+1 | [20,19, 18,17, 24] | 5.7380 | 15.6437 | 21.3817 | [1.0000] | [0.2085] | [0.2025, 0.1999, 0.1966, 0.1925] |
| 4+2 | [20,19, 18,17, 24,22] | 6.3829 | 14.7948 | 21.1777 | [0.5193, 0.4807] | [0.1686, 0.1561] | [0.1734, 0.1708, 0.1676, 0.1635] |
| 4+3 | [20,19, 18, 17, 24,22,21] | 6.3829 | 14.7948 | 21.1777 | [0.5193, 0.4807, 0] | [0.1686, 0.1561, 0] | [0.1734, 0.1708, 0.1676, 0.1635] |
| 4+4 | [20,19, 18,17, 24,22, 21,20] | 6.3829 | 14.7948 | 21.1777 | [0.5193, 0.4807, 0, 0] | [0.1686, 0.1561, 0, 0] | [0.1734, 0.1708, 0.1676, 0.1635] |

We can also adapt Algorithm 2 to incorporate the direct beams. We first compare different number of reflected beams to include under the uniform $\boldsymbol{r}$ (but not uniform $\boldsymbol{q}$) approximation to find the optimal reflection rank $rk^*$, then compute the optimal resource allocation $(\boldsymbol{q}_{\boldsymbol{D}}^*, \boldsymbol{q}_{\boldsymbol{R}}^*, \boldsymbol{r}^*)$ with *Opt_Dir_RIS_rank*() under $rk^*$. The result is Algorithm 4.

Algorithms 1-4 can be used for point-to-point link or for multiple users (MU-MIMO) in the down link. All the algorithms are very efficient. Operating within the singular space of the channel, the size of the problem only depends on the number of distinct beams, not on the number of RIS elements on the reflection surface. Our experiments showed these algorithms converge very swiftly, making them applicable to large RIS networks.

## 5. Conclusion

We have developed a framework for RIS-assisted MIMO communication in sparse channel typically found at high frequency range such as mmW and THz. By exploring the sparse property and the high angular resolution of large antenna arrays, we decompose the channel into a set of mutually orthogonal singular modes and manipulate these modes directly. We derived the capacities of the RIS reflection channel and developed efficient algorithms to achieve them. We also compared the traditional MIMO channel and the RIS reflection channel. Our result cautions against using RIS for high order reflection.

## Appendix A: Proof of Theorem 1

For SVD of a matrix $\boldsymbol{H} = \boldsymbol{U}\boldsymbol{S}\boldsymbol{V}^{\boldsymbol{H}}$, a sufficient condition for $\boldsymbol{V} = \boldsymbol{I}$ is:

$$\boldsymbol{H}^{\boldsymbol{H}}\boldsymbol{H} = \boldsymbol{V}\boldsymbol{S}^{\boldsymbol{H}}\boldsymbol{U}^{\boldsymbol{H}}\boldsymbol{U}\boldsymbol{S}\boldsymbol{V}^{\boldsymbol{H}} = \boldsymbol{V}\boldsymbol{S}^{\boldsymbol{H}}\boldsymbol{S}\boldsymbol{V}^{\boldsymbol{H}} = \boldsymbol{S}^{\boldsymbol{H}}\boldsymbol{S} = diag([s_1^2, s_1^2, \dots, s_N^2]). \quad \text{(A.1)}$$

For $\boldsymbol{H}^{\boldsymbol{H}}\boldsymbol{H}$ to be diagonal, the columns of $\boldsymbol{H}$ need to be orthogonal to each other, i.e. $\langle h_i^c, h_j^c \rangle = 0$ for $i \neq j$. This condition is satisfied when there is at most one non-zero element in each row, therefore $\boldsymbol{V} = \boldsymbol{I}$. The singular values of $\boldsymbol{H}$ are (without ordering)

$$s_i^2 = \langle h_i^c, h_i^c \rangle, \; s_i = |h_i^c|. \quad \text{(A.2)}$$

Q.E.D.

**Appendix B: The Algorithms**

**Beam pairing algorithm:**

Input: TX-RIS channel $\boldsymbol{H_{v1}}$, RIS-RX channel $\boldsymbol{H_{v2}}$. Output: Pairs of TX-RIS beam and RIS-RX beam $\left(i_1^j, k_2^j\right)$ in descending order with respect to effective gain $g_j$, and the phase control parameter $N_c^j$ of the corresponding subarray.

Sort the AOAs ($\{i_1\}$) of $\boldsymbol{H_{v1}}$ in descending order w.r.t. $|\boldsymbol{H}_{v1}(i_1, k_1)|$. Sort the AODs ($\{k_2\}$) of $\boldsymbol{H_{v2}}$ in descending order w.r.t. $|\boldsymbol{H}_{v2}(:, k_2)|$. The *j*-th beam pair comprises of $\left(i_1^j, k_2^j\right)$. The effective gain is $g_j = \left|\boldsymbol{H}_{v2}(:, k_2^j)\right|^2 \left|\boldsymbol{H}_{v1}(i_1^j, k_1^j)\right|^2$. The corresponding subarray has phase control parameter $N_c^j$: $N_c^j = k_2^j - i_1^j$.

**Joint TX power-RIS allocation Algorithm 1 (for reflection channel only)**:

Input: A set of beam pairs generated by the beam pairing algorithm with normalized ***snr*** sorted in descending order. Output: the optimal allocation ($\boldsymbol{q}^*$,$\boldsymbol{r}^*$) and the maximal capacity $c^*$.

```
function (q*, r*, c*) = Joint_power_RIS_alloc_1(snr) {
    J_max = length(snr); rk = J_max;
    for k=1: J_max {
        (Q(k,:), R(k,:), c(k)) = Opt_RIS_rank(snr(1:k));
        If c(k) ≤ c(k − 1) {
            rk = k − 1;    break;
        }
    }
    q*= Q(rk,:);  r*=R(rk,:); c* = c(rk);
```

}

function ($\boldsymbol{q}^*, \boldsymbol{r}^*, c^*$) = *Opt_RIS_rank*($\boldsymbol{snr}$) {

$N = length(\boldsymbol{snr});\ \boldsymbol{r}^{(0)} = \frac{ones(N,1)}{K};$

$i = 1;$

While ($i \leq I_{max}$) {

$\boldsymbol{r}^{(i)} = Waterfill\left(\frac{\mathbf{1}}{power(\boldsymbol{r}^{(i-1)},2).*\boldsymbol{snr}}, 1\right);$

if ($\left|\boldsymbol{r}^{(i)} - \boldsymbol{r}^{(i-1)}\right| < \epsilon_{conv}$) {

$(\boldsymbol{q}^*, \boldsymbol{r}^*) = (\boldsymbol{r}^{(i)}, \boldsymbol{r}^{(i)});$ break;

}

$i = i + 1;$

}

$c^* = \sum_{k=1}^{N} log_2(1 + \boldsymbol{r}^*(k)^3 \boldsymbol{snr}(k));$

}

function $\boldsymbol{p} = Waterfill(\boldsymbol{n}, P_{tot})$ is the standard water filling algorithm for parallel AWGN channels, where $\boldsymbol{n}$ is the noise power vector, $P_{tot}$ is total power, and $v$ is the water level:

$$p_i = (v - n_i)^+, \sum_i p_i = P_{tot}.$$

**Joint TX power-RIS allocation Algorithm 2 (for reflection channel only)**:

Input and output: same as Algorithm 1

function ($\boldsymbol{q}^*, \boldsymbol{r}^*, c^*$) = *Joint_power_RIS_alloc_2*($\boldsymbol{snr}$) {

$J_{max} = length(\boldsymbol{snr}); rk = J_{max};$

for $k = 1: J_{max}$ {

$\boldsymbol{c}(k)= \sum_{i=1}^{k} log_2(1 + \boldsymbol{snr}(i)/k^3)$ ;

If $c(k) \leq c(k-1)$ {

$rk = k - 1;$ break;

}

}

$(\boldsymbol{q}^*, \boldsymbol{r}^*, c^*)$ = *Opt_RIS_rank*($\boldsymbol{snr}(1: rk)$);

}

**Joint TX power-RIS allocation Algorithm 3 (for reflection & direct channel)**:

Input: A set of direct beams with $\boldsymbol{snr_D}$, a set of reflected beam pairs generated by the beam pairing algorithm with $\boldsymbol{snr_R}$ sorted in descending order.

Output: the optimal allocation $(\boldsymbol{q_D^*}, \boldsymbol{q_R^*}, \boldsymbol{r}^*)$ and the maximal capacity $c^*$.

function $(\boldsymbol{q_D^*}, \boldsymbol{q_R^*}, \boldsymbol{r}^*, c^*)$ = *Joint_power_Dir_RIS_alloc_1*($\boldsymbol{snr_D}, \boldsymbol{snr_R}$) {

$J_{max} = length(\boldsymbol{snr_R}); rk = J_{max};$

for $k$=1:$N_R$ {

$(\boldsymbol{q_D}(k), \boldsymbol{q_R}(k), \boldsymbol{r}(k), c(k))$ = *Opt_Dir_RIS_rank*($\boldsymbol{snr_D}, \boldsymbol{snr_R}(1: k)$);

If $c(k) \leq c(k-1)$ {

$rk = k - 1;$ break;

}

}

$\boldsymbol{q_D^*} = \boldsymbol{q_D}(rk);\ \boldsymbol{q_R^*} = \boldsymbol{q_R}(rk); \boldsymbol{r}^* = \boldsymbol{r}(rk);\ c^* = c(rk);$

}

function $(\boldsymbol{q_D^*}, \boldsymbol{q_R^*}, \boldsymbol{r}^*, c^*)$ = *Opt_Dir_RIS_rank*($\boldsymbol{snr_D}, \boldsymbol{snr_R}$) {

$N_D = length(\boldsymbol{snr_D}); N_R = length(\boldsymbol{snr_R});$

$\boldsymbol{r}^{(0)} = \frac{ones(1,N_R)}{N_R}$; $\boldsymbol{q_R}^{(0)} = \frac{ones(1,N_R)}{N_D+N_R}$; $\boldsymbol{q_D}^{(0)} = \frac{ones(1,N_D)}{N_D+N_R}$;

$i = 1;$

While ($i \leq I_{max}$) {

$\boldsymbol{q}^{(i)} = Waterfill\left(\frac{\mathbf{1}}{[power(\boldsymbol{r}^{(i-1)},2).*\boldsymbol{snr_R},\ \boldsymbol{snr_D}]}, 1\right)$**;**

$\boldsymbol{q_R}^{(i)} = \boldsymbol{q}^{(i)}(1{:}\,N_R)$; $\boldsymbol{q_D}^{(i)} = \boldsymbol{q}^{(i)}(N_R + 1{:}\,N_R + N_D)$;

$\boldsymbol{r}^{(i)} = \frac{\boldsymbol{q_R}^{(i)}}{sum(\boldsymbol{q_R}^{(i)})}$**;**

if ($\left|\boldsymbol{q}^{(i)} - \boldsymbol{q}^{(i-1)}\right| < \epsilon_{conv}$) {

$\boldsymbol{r}^* = \boldsymbol{r}^{(i)}$; $\boldsymbol{q}_{\boldsymbol{R}}^{*} = \boldsymbol{q}_{\boldsymbol{R}}^{(\boldsymbol{i})}$; $\boldsymbol{q}_{\boldsymbol{D}}^{*} = \boldsymbol{q}_{\boldsymbol{D}}^{(\boldsymbol{i})}$;

break;

}

$i = i + 1;$

}

$$c^* = \sum_{k=1}^{N_R} log_2(1 + \boldsymbol{q}_{\boldsymbol{R}}^{*}(k)\boldsymbol{r}^*(k)^{\mathbf{2}}\boldsymbol{snr_R}(k))$$

$$+ \sum_{k=1}^{N_D} log_2(1 + \boldsymbol{q}_{\boldsymbol{D}}^{*}(k)\boldsymbol{snr_D}(k))\ ;$$

}

**Joint TX power-RIS allocation Algorithm 4 (for reflection & direct channel)**:

Input and output: same as Algorithm 3

function ($\boldsymbol{q}_{\boldsymbol{D}}^{*}, \boldsymbol{q}_{\boldsymbol{R}}^{*}, \boldsymbol{r}^*, c^*$) = *Joint_power_Dir_RIS_alloc_2*($\boldsymbol{snr_D}, \boldsymbol{snr_R}$) {

$N_D = length(\boldsymbol{snr_D}); N_R = length(\boldsymbol{snr_R});$

$rk = N_R;$

for $k$=1: $N_R$ {

$\boldsymbol{Q}(k,:) = Waterfill\left(\frac{\mathbf{1}}{[\boldsymbol{snr_R}(1{:}k)/k^2,\ \boldsymbol{snr_D}]}, 1\right)$;

$\boldsymbol{q_R}(k,:) = \boldsymbol{Q}(k, 1{:}N_R)$; $\boldsymbol{q_D}(k,:) = \boldsymbol{Q}(k, N_R + 1{:}N_R + N_D)$;

$$c(k) = \sum_{i=1}^{N_R} log_2(1 + \frac{\boldsymbol{q_R}(k,i)\boldsymbol{snr_R}(i)}{k^2}) + \sum_{i=1}^{N_D} log_2(1 + \boldsymbol{q_D}(k,i)\boldsymbol{snr_D}(i));$$

If $c(k) \leq c(k-1)$ {

$rk = k - 1;$ break;

}

}

$(\boldsymbol{q_D^*}, \boldsymbol{q_R^*}, \boldsymbol{r}^*, c^*)$ = *Opt_Dir_RIS_rank*($\boldsymbol{snr_D}$, $\boldsymbol{snr_R}(1{:}rk)$);

}